\documentclass[runningheads]{llncs}
\usepackage{graphicx}
\usepackage{booktabs} 
\usepackage{cite}
\usepackage{amsfonts, amssymb}
\usepackage{amsmath}
\usepackage{algorithm}
\usepackage{algorithmic}
\usepackage{threeparttable}
\usepackage{marvosym}

\begin{document}

\title{Spatiotemporal Graph Neural Network Modelling Perfusion MRI}
\titlerunning{Spatiotemporal GNN Modelling Perfusion MRI}
\author{Ruodan Yan\inst{1} \and
Carola-Bibiane Schönlieb\inst{1} \and
Chao Li\inst{1,2}\textsuperscript{\Letter}}
\authorrunning{R. Yan et al.}
\institute{Department of Applied Mathematics and Theoretical Physics, University of Cambridge
\and
Department of Clinical Neurosciences, University of Cambridge
\\
\email{cl647@cam.ac.uk}}
\maketitle              % typeset the header of the contribution
\begin{abstract}
Perfusion MRI (pMRI) offers valuable insights into tumor vascularity and promises to predict tumor genotypes, thus benefiting prognosis for glioma patients, yet effective models tailored to 4D pMRI are still lacking. This study presents the first attempt to model 4D pMRI using a GNN-based spatiotemporal model (\textbf{PerfGAT}), integrating spatial information and temporal kinetics to predict Isocitrate DeHydrogenase (IDH) mutation status in glioma patients. Specifically, we propose a graph structure learning approach based on edge attention and negative graphs to optimize temporal correlations modeling. Moreover, we design a dual-attention feature fusion module to integrate spatiotemporal features while addressing tumor-related brain regions. Further, we develop a class-balanced augmentation methods tailored to spatiotemporal data, which could mitigate the common label imbalance issue in clinical datasets. Our experimental results demonstrate that the proposed method outperforms other state-of-the-art approaches, promising to model pMRI effectively for patient characterization.

\end{abstract}
\section{Introduction}
Glioma is the most common adult malignant brain tumor, characterized by dismal prognosis\cite{louis_2021_2021}. Isocitrate DeHydrogenase (IDH) mutation is one of the most important genetic markers for gliomas. Clinically, determining IDH mutation status relies on invasive procedures, i.e., immunohistochemistry and gene sequencing of surgical samples, which are often time-consuming and less feasible for some patients. In parallel, MRI offers a non-invasive characterization of brain tumors. Previous MRI  studies\cite{wei_multi-modal_2023, wang_multi-task_2023} show that the genotype of glioma can be predicted using machine learning, offering valuable clinical decision support.

Perfusion MRI (pMRI) is a technique to reflect brain vascularity and microcirculation. By capturing the dynamics of contrast agents passing through the vasculature, pMRI acquires sensitive spatiotemporal information to probe neovascularization caused by angiogenesis, which could indicate tumor aggressiveness and better predict the genotype of glioma over conventional MRI. However, it remains a challenge to develop effective spatiotemporal modelling for pMRI.

Previous studies predominantly rely on pre-defined pharmacokinetic models\cite{li_decoding_2019}, leveraging mathematical representations to calculate several perfusion parameters from the kinetic curve, e.g., relative cerebral blood volume (rCBV). However, the derived parametric maps are inherently simplified and limited in modelling complex physiological kinetics. Although various attempts have been made to develop machine learning/deep learning approaches\cite{Pinto_2018,heo_deep_2023} for extracting more comprehensive features from parametric maps, sophisticated spatiotemporal modelling of pMRI remains an unmet need to decode perfusion kinetics effectively.

% However, as these models are trained purely by CT intensities, the crucial information about brain structure and function embedded in the pMRI cannot be included. 

%these models are typically designed for CT, which have better resolution than pMRI. This may exacerbate the risk of overfitting in pMRI analysis due to the large number of parameters inherent in these models. Additionally, image-based methods struggle to capture the intricate interactions between different brain regions constructing in non-Euclidean space, which is crucial to better understand glioma invasion. Moreover, the computational complexity and memory requirements of these models may limit their practical application in pMRI analysis.

%it is central to integrate spatial and temporal features in order to characterize vascular networks based on regional blood and perfusion kinetics. 
Scanty work developed spatiotemporal models to learn kinetics in perfusion imaging. Zhang et al.\cite{yao_deepprognosis_2021} proposed 3D-ResNet based ConvLSTM\cite{shi_convolutional_2015} framework, to predict patient survival. However, as this model is based on contrast-enhanced CT, it cannot include detailed tissue structure and regional perfusion available from pMRI. Another spatiotemporal model, SwiFT\cite{kim2023swift}, is proposed for functional MRI (fMRI) based on the Swin Transformer\cite{liu2021swin}. This model employs a window attention mechanism using the image patches defined by the Euclidean space, which limits its ability to model the perfusion patterns of anatomical  regions. 

Brain networks promise to comprehensively characterise the global and regional brain, where brain regions from the prior atlas are treated as nodes and their connectivities as edges\cite{wei_quantifying_2021}. Based on fMRI, graph neural networks (GNNs)  have been shown to characterize brain function and disorders through modelling brain networks\cite{li_braingnn_2021}. Particularly, GNNs can model the spatiotemporal graphs, e.g., STGCN for traffic forecasting\cite{yu_spatio-temporal_2018}, which promise to model the kinetics of perfusion networks. However, existing spatiotemporal graph models are not designed for medical images and, therefore, may be unable to capture the complex spatial structures and intricate physiological processes from the perfusion graph. 
% and DySAT for representation learning\cite{sankar_dynamic_2019}

Further, both function and perfusion networks typically have noisy, dense connections, leading to overfitting risk in GNN modelling. Graph structure learning promises to refine the connectivities in functional graphs\cite{zhang_brainusl_2023}. However, this method may be less effective in perfusion graphs due to the intrinsic differences between functional and perfusion graphs in topological properties and temporal resolution of raw MRI. It remains an unmet need to develop approaches to optimize the structure of perfusion graphs reflecting vasculature networks.

In this study, we propose a spatiotemporal GNN framework to learn the kinetics of pMRI and predict the genotype of glioma. Our contributions include:
\begin{enumerate}
    \item\textbf{Perfusion Graph Structure Learning.} We propose a graph structural learning module utilizing edge attention and a negative graph to optimize the temporal graph structure. Specifically, edge attention allows the model to adaptively assign attention to different edges based on their importance in capturing temporal correlations and identifying the most relevant graph structure. Moreover, a negative graph provides complementary information to identify potential missing important connections.
    \item\textbf{Dual-attention Guided Spatiotemporal Feature Fusion.} To optimize the encoder's capability in extracting pertinent spatiotemporal features while mitigating the influence of concurrent pathologies, we introduce a dual-attention mechanism, which systematically prioritizes brain regions by node attention and fuse spatiotemporal features with semantic attention, enabling the identification of most tumor-specific spatiotemporal features.
    \item\textbf{Class-balanced feature augmentation}. Class imbalance is a significant challenge in predicting genotypes due to natural incidence. Common methods for mitigating class imbalance, such as oversampling and cost-sensitive learning, are limited by overfitting and introducing noise or distortion to data distribution\cite{Chawla_2002, cui2019classbalanced}. Here we propose a spatiotemoral augmentation method which recombines local and graph features from the minority class during the retraining stage, and incorporates a class-balanced classifier retraining approach\cite{kang2020decoupling}. Thereby improving model generalizability and robustness.
\end{enumerate}
As far as we know, this is the first work to develop spatiotemporal graph learning for characterizing pMRI. Our experimental results show that our method outperforms other state-of-the-art models in predicting the genotype of glioma. 

\begin{figure}
\includegraphics[width=\textwidth]{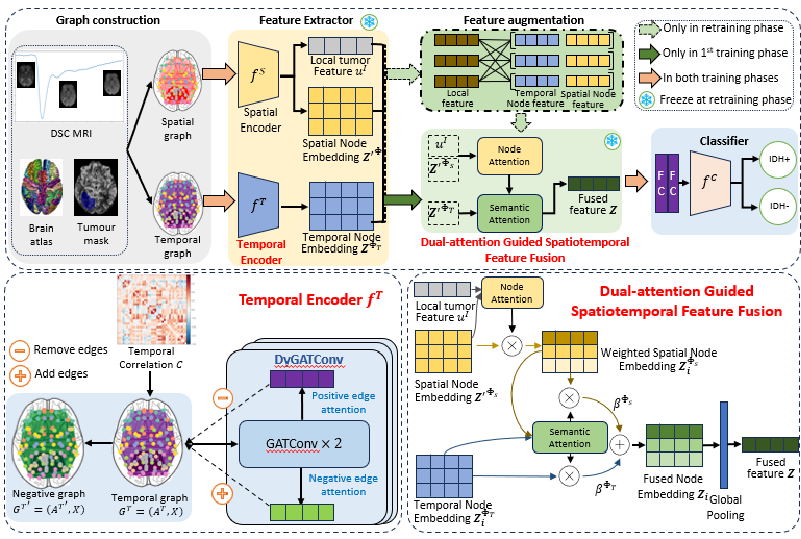}
\caption{Model Design. \textit{Top}. Training Pipeline. DSC-MRI data, brain atlas and tumor masks construct input graphs. 
Features of focal tumor $u^I$, spatial graph $\mathbf{Z'}^{\Phi_S}$, and temporal graph $\mathbf{Z'}^{\Phi_T}$ are extracted by encoders. During retraining, features are augmented before dual-attention fusion. Resultant fused feature $\mathbf{Z}$ is used for IDH prediction. 
\textit{Bottom left}. Temporal encoder. Pairs of $G^T$ and its negative graph $G^{T'}$ are fed into the DyGATConv module for iterative connection adjustment based on the edge attention. \textit{Bottom right}. Dual-attention Feature Fusion. $u^I$ and $\mathbf{Z'}^{\Phi_S}$ are fused using node attention. The resulting weighted spatial node embedding is then fused with $\mathbf{Z'}^{\Phi_T}$ using semantic attention, resulting in fused feature $\mathbf{Z}$ by global pooling.} \label{overview}
\end{figure}

\section{Methodology}

\subsubsection{Overview.} The proposed PerfGAT (Fig.~\ref{overview}) firstly constructs a spatiotemporal graph using DSC-MRI, brain atlas and tumor masks to incorporate focal tumors. Subsequently, spatial and temporal graph features and local tumor features are generated using distinct encoders. To reduce noisy temporal connections, we employ a graph structure learning approach based on edge attentions (details in~\ref{gsl}). The dual-attention feature fusion mechanism (details in~\ref{dual_attn}) strategically integrates features across local tumor and spatial and temporal graphs, yielding comprehensive representations from pMRI. Finally, to tackle the class-imbalance issue while avoiding data distortion, we employ a recombining augmentation mechanism tailored to spatiotemporal graph (details in~\ref{aug}). 

%(1) Graph structure learning based on edge attention mechanism and negative graph in temporal encoder; (2) Dual-attention guided spatiotemporal feature fusion to fuse local tumor, spatial graph and temporal graph sequentially by node attention and semantic attention; and (3) Class-balanced retraining strategy using spatiotemporal feature augmentation mechanism recombining local and global (graph) features when resampling.

\subsubsection{Spatiotemproal graph generation.}

We construct a spatiotemporal graph $\mathbf{G}$ from the raw DSC-MRI using the SRI24/TZO atlas\cite{rohlfing_sri24_2010} and tumor masks. Specifically, the nodes are defined as the brain regions of the atlas and the tumor region is segmented based on the tumor mask and taken as a separate node. The average time-signal curve of brain regions is extracted as node features $\mathbf{X}^B\in \mathbb{R}^{N\times T}$, where N is the node number, and T is the time length. 

The perfusion graph consists of temporal and spatial connections. 1) Temporal connections represent the temporal correlations of perfusion kinetics across brain regions. We first derive a temporal correlation matrix $\mathbf{C}\in \mathbb{R}^{N\times N}$, where $C_{ij}$ represent the correlation coefficient between node pairs. Thresholding this matrix yields the temporal adjacency matrix $\mathbf{A}^T\in \mathbb{R}^{N\times N}$, where $\mathbf{A}^T_{ij}=1$ if $C_{ij}$ exceeds a predefined threshold, and $0$ otherwise. 2) Spatial connections are defined by the Euclidean distance of brain regions. Each node is connected to its k nearest neighbours, forming spatial adjacency matrix $\mathbf{A}^S\in \mathbb{R}^{N\times N}$, where $A^S_{ij}=1$ indicates connected edges and $A^S_{ij}=0$ indicates unconnected edges. This results in temporal graphs $\mathbf{G}^T=(\mathbf{X}^B,\mathbf{A}^T)$ and spatial graphs $\mathbf{G}^S=(\mathbf{X}^B,\mathbf{A}^S)$. 

\subsection{Graph Structure Learning}\label{gsl}
\begin{algorithm}
\renewcommand{\algorithmicrequire}{\textbf{Input:}}
\renewcommand{\algorithmicensure}{\textbf{Output:}}
\caption{Graph Structure Learning}\label{GSL}
\begin{algorithmic}
\REQUIRE $G^T = (X^B, A^T), \alpha, \beta, l \gets 0$
\REPEAT
\STATE $l \gets l+1$
\STATE $G^{T^-} = (X^B, A^{T^-}) \gets \text{Negative\_graph} (G^T)$
\STATE Update $E^+$ from $G^T$ based on Equation~(\ref{GAT})
\STATE Update $E^-$ from $G^{T^-}$ based on Equation~(\ref{GAT})
\STATE $\text{edges\_to\_delete} \gets \text{indices of lowest $\alpha$ }  e^+_{ij} \text{ in } E^+$
\STATE $\text{edges\_to\_add} \gets \text{indices of highest $\beta$ }  e^-_{ij} \text{ in } E^-$
\STATE $G^T \gets \text{remove\_edges}(\text{edges\_to\_delete})$
\STATE $G^T \gets \text{add\_edges}(\text{edges\_to\_add})$
\STATE Update $X^B$ based on Equation~(\ref{GAT})
\UNTIL $l = max\_layer$
\RETURN $G^T$
\end{algorithmic}
\end{algorithm}
Our approach to graph structure learning is aimed at optimizing temporal graph connectivity to mitigate noisy temporal connections and enhance adaptability to task-specific nuances (Alg.~\ref{GSL}). The graph structure learning module operates within each layer to dynamically adjust the graph structure. It begins by generating a negative graph, denoted as $G^{T^-}=(A^{T^-}, X^B)$, capturing the edges absent in the original graph while maintaining the same nodes. The edge attentions of the original graph and the negative graph are computed through the graph attention layer (GATConv\cite{veličković2018graph}) based on Eq.\ref{GAT}, where $e^+_{ij}$ and $e^-_{ij}$ denote the attention scores for the edges in each graph, respectively. Subsequently, edges with low $e^+_{ij}$ are considered noisy edges to remove. Conversely, edges with high $e^-_{ij}$ representing informative connections absent in the original graph are added.
% \begin{equation}
%     A^{T^{(l+1)}}_{ij} = \alpha(\mathbf{W}_A*[x_i\Vert A^{T^{(l)}}_{ij}\Vert x_j])
% \label{edge}
% \end{equation}
\begin{equation}
     e_{ij}=\textit{softmax}_i(\textit{LeakyReLu}(\Vec{w}_a^T[\mathbf{W}_A*x_i\Vert \mathbf{W}_A*x_j]))
\label{GAT}
\end{equation}
where $\mathbf{W}_A$ is a weight matrix; $x_i$ and $x_j$ are node feature $i$ and $j$; $\Vec{w}_a$ is a weight vector; $e_{ij}$ is edge attention score of  connecting nodes $i$ and $j$.

After acquiring the edge attention, the node feature is updated by Eq.\ref{node_feat}.
\begin{equation}
x^{(l+1)}_i=\sum_{j\in \mathcal{N}(i)}e_{ij}*\mathbf{W}_n^{(l)}*x^{(l)}_j\label{node_feat}
\end{equation}
where $x^{(l+1)}_i$ is the feature of node $i$ in layer $(l+1)$; $x^{(l)}_j$ is the feature of node $j$ in layer $l$; $e_{ij}$ is edge attention score connecting node $i$, $j$; $\mathbf{W}_n$ is a weight matrix.

\subsection{Dual-attention guided spatiotemporal feature fusion}\label{dual_attn}
In parallel to temporal encoder ($f^T(\cdot)$),  spatial graph is encoded by GAT layers ($f^\text{GAT}(\cdot)$) and local tumor is encoded by 3DResNet + ConvLSTM\cite{yao_deepprognosis_2021} ($f^I(\cdot)$). Embeddings of temporal node  $\mathbf{Z}^{\Phi_T}=f^T (X^B, A^T)$, spatial node $\mathbf{Z}^{\Phi'_S}=f^\text{GAT} (X^B, A^S)$, and local tumor $u^I=f^I(X^I)$ ($\mathbf{X}^I$: local tumor image) are acquired.

The proposed dual-attention mechanism is structured by node attention and semantic attention. It first focuses on the spatial features via node attention to emphasize the tumor-related node embedding, consequently mitigating the influence of noise inherent in complex high-dimensional perfusion data and confounding factors. After project spatial node embeddings and tumor features into the shared latent space, The node attention is computed as follows:
\begin{equation}
\begin{array}{cc}
     a_i=S(f^\text{MLP}_N(z^{\Phi'_S}_i),f^\text{MLP}_F(u^I))&
\end{array}
\end{equation}

where $a_i$ denotes the attention of the $i$th node; $z^{\Phi'_S}_i$ is the spatial feature of $i$th node; $S(\cdot)$ is the cosine similarity function, and $f^\text{MLP}_N(\cdot)$ and $f^\text{MLP}_F(\cdot)$ are projection heads for node embeddings and tumor features.

Subsequently, inspired by the HAN model\cite{wang2021heterogeneous}, we leverage semantic attention to further fuse spatial and temporal node embeddings. The spatial and temporal node embeddings are taken as input and projected to a latent space ($f^\text{MLP}(\cdot)$). The transformed embeddings are measured using a semantic-level attention vector $\Vec{q}$. We then average all the node embeddings as the significance of each dimension (spatial or temporal). The learned weights of each type $\beta_{\Phi_d} (d\in (S, T))$ are then obtained by softmax normalization to compute the final node embeddings $\mathbf{Z}$:
\begin{equation}
    \beta_{\Phi_d}=\textit{Softmax}(\frac{1}{\lvert \mathcal{N}\rvert}\sum_{i\in \mathcal{N}}\Vec{q}\cdot \textit{tanh}(f^\text{MLP}(\mathbf{Z}^{\Phi_d})))
\end{equation}
\begin{equation}
     % \beta_{\Phi_d}=\text{Softmax}(w^{\Phi_d})&  \\
     \mathbf{Z}=\beta^{\Phi_S}\cdot\mathbf{Z}^{\Phi_S} + \beta^{\Phi_T}\cdot\mathbf{Z}^{\Phi_T}
\end{equation}
% \begin{equation}
%     (\beta_{\Phi_T}, \beta_{\Phi_S})=\text{att}_\text{sem}(\mathbf{Z}^{\Phi_T}, \mathbf{Z}^{\Phi_S})
% \end{equation}

%The utilization of attention during the training of the brain network encoder allows for the extraction of features most relevant to the tumor from the brain network,  The feature extraction of the brain networks is defined as $u_i^B=P(f^B (X_i^B ),u_i^I )$, where $X_i^B$, $u_i^B$ and $u_i^I$ are the node features, graph features and tumor features of the $i$th patient, respectively. $f^B$ is the EGAT-based brain network encoder and $P(\cdot)$ is the global pooling layer responsible for generating graph-level features.
\begin{table}
\centering
\caption{The results for IDH classification on the meta-cohort dataset.}\label{baseline}
\resizebox{\linewidth}{!}{
\begin{tabular}{c c c c c c c}
\toprule
Model Name &  ACC ($\%$) & B-ACC ($\%$) & SPEC ($\%$) & SENS ($\%$) & F1 ($\%$) & AUC ($\%$)\\
\midrule
T1+CNN & \textbf{92.5} & 50.0 & \textbf{100} & 00.0 & 00.0 & 78.4\\
rCBV+CNN & \textbf{92.5} & 50.0 & \textbf{100} & 00.0 & 00.0 & 76.5\\
\midrule
ResNet + ConvLSTM\cite{shi_convolutional_2015} & 80.6 & 71.1 & 82.3 & 60.0 & 31.6 & 81.0\\
SwiFT\cite{kim2023swift} & 77.6 & 69.5 & 79.0 & 60.0 & 19.5 & 80.0\\    % done
% ResNet + CNN + transformer & -& - & - & - & -\\
\midrule
GCN\cite{kipf2017semisupervised} & 57.3 & 69.4 & 55.4 & \textbf{83.3} & 20.8 & 77.5\\
GIN\cite{xu2019powerful} & 92.1 & 57.1 & 97.6 & 16.7 & 22.2 & 57.4\\
GAT\cite{veličković2018graph} & 69.7 & 68.3 & 69.9 & 66.7 & 22.9 & 75.1\\
EGAT\cite{kaminski_rossmann-toolbox_2022} &  65.2 & 65.9 & 65.1 & 66.7 & 20.5 & 77.1\\
STGCN\cite{yu_spatio-temporal_2018} & 52.2 & 55.8 & 51.6 & 60.0 & 15.8 & 63.2\\
% DySAT\cite{sankar_dynamic_2019} & - & - & - & - & -\\
BrainGNN\cite{li_braingnn_2021} & 85.1 & 55.2 & 90.3 & 20.0 & 16.7 & 58.7\\
%MVS-GCN\cite{wen_mvs-gcn_2022} & - & - & - & - & - & -\\
\midrule
PerfGAT &  88.8 & \textbf{86.2} & 89.2 & \textbf{83.3} & \textbf{50.0} & \textbf{94.4}\\
\bottomrule
\end{tabular}
}
\end{table}

\subsection{Class-balanced Recombining Augmentation}\label{aug}
To address the limitations of random resampling methods and enhance the robustness of our spatiotemporal GNN model, we propose a class-balanced recombining augmentation technique tailored for multimodal graph learning. Here, we randomly select pairs of minority class samples $h_\text{minority}=(u^I, u^B)$ where $u^B=(Z^{\Phi_S}, Z^{\Phi_T})$, and $h'_\text{minority}=(u^{I'}, u^{B'})$. By recombining features within the minority class, we generate augmented samples $h_\text{aug1}=(u^I, u^{B'})$ and $h_\text{aug2}=(u^{I'}, u^B)$. This augmentation process ensures the synthesis of sufficient minority samples to achieve class balance within a highly biased dataset. By encouraging the model to learn more diverse representations of the minority class, our innovative recombining method significantly enhances its generalizability to unseen data, further strengthening the effectiveness of our spatiotemporal GNN model.

%Here, we design a classifier retraining approach \cite{kang_decoupling_2019}, where they stated that instead of performing class-balanced resampling before training, it is more useful to first train the model without resampling, then freeze the feature extractor and retrain the classifier with class-balanced resampling. They proved that the ability of feature extractor is less affected by the class imbalance problem, while the training of classifier is more essential to obtain better decision boundaries.

\begin{figure}
\includegraphics[width=\textwidth]{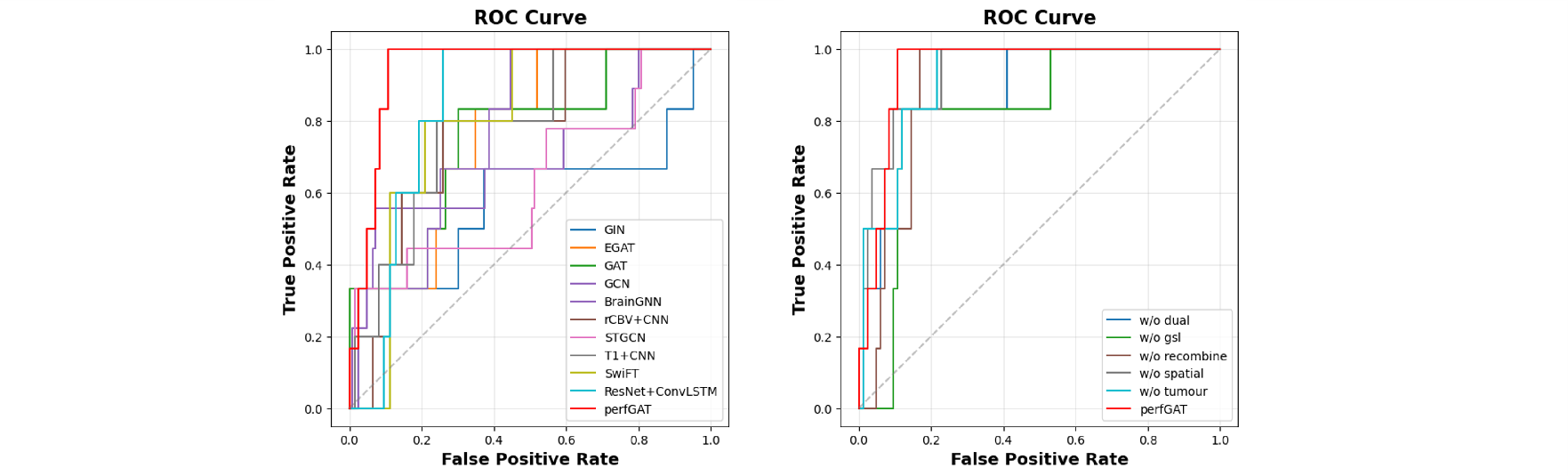}
\caption{ROC curves of comparison models (left), and ablation models (right)} \label{ROC}
\end{figure}

\section{Experiments and Results}

\subsection{Datasets and Image Pre-processing}
We compiled a meta-cohort with Dynamic Susceptibility Contrast (DSC) MRI of 444 glioma patients, including the UPenn-GBM dataset available from The Cancer Imaging Archive (TCIA), and an in-house dataset. Among patients, 30 (6.8$\%$) are IDH mutants, while 414 (93.2$\%$) are wildtype IDH. Patients were divided into training, validation, and testing sets in a 7:1:2 ratio.

The DSC images were first co-registered with post-contrast T1 using the FMRIB Linear Image Registration Tool (FLIRT) from FSL. Subsequently, skull stripping was performed using the Brain Extraction Tool (BET) in FSL. Normalization included histogram matching and voxel smoothing with SUSAN noise reduction according to the BraTS preprocessing procedures. Finally, all MRIs were non-linearly transformed to the standard space by co-registering them to the SRI atlas \cite{rohlfing_sri24_2010}, using the Advanced Normalization Tools (ANTs). 

Tumor masks, delineating the tumor core, were generated following standard procedures outlined in the brain tumor segmentation benchmark. Initially, an initial tumor segmentation (auto-mask) was generated using a pre-trained nnUNet, which was then manually inspected and corrected by two clinical experts.

% The rCBV maps were derived using Cancer Imaging Phenomics Toolkit (CaPTk). 

\subsection{Experimental Settings and Implementations}
We used two 24GB NVIDIA RTXA5000 GPU and PyTorch.The main training stage had a maximum of 200 epochs, and an early stopping was used to avoid over-fitting when validation loss no longer decreased over 10 epochs. For retraining, we set the number of epochs as 20. The batch size for graph and 3D image data was 10, while for 4D image data was 1 to reduce computational costs. The optimizer was Adam\cite{kingma2017adam}, and the learning rate was 1e-4 for all trainings. 

\begin{table}
\centering
\caption{The ablation results for IDH classification on the meta-cohort dataset.}
\resizebox{\linewidth}{!}{
\begin{threeparttable}
\begin{tabular}{c c c c c c c}
\toprule
Model Config &  ACC ($\%$) & B-ACC ($\%$) & SPEC ($\%$) & SENS ($\%$) & F1 ($\%$) & AUC ($\%$)\\
\midrule
w/o Graph structure learning & 86.5 & 77.3 & 88.0 & 66.7 & 40.0 & 82.3\\
w/o Dual-attention & \textbf{88.8} & 78.5 & \textbf{90.4} & 66.7 & 44.4 & 88.8\\
w/o Recombine augmentation & 83.1 & 83.2 & 83.1 & \textbf{83.3} & 40.0 & 89.4\\
w/o Spatial graph* & 77.5 & 80.2 & 77.1 & \textbf{83.3} & 33.3 & 93.2\\
w/o Tumor local** & 78.7 & 80.8 & 78.3 & \textbf{83.3} & 34.5 & 92.0\\
\midrule
PerfGAT &  \textbf{88.8} & \textbf{86.2} & 89.2 & \textbf{83.3} & \textbf{50.0} & \textbf{94.4}\\
\bottomrule
\end{tabular}
\begin{tablenotes}
\item *Only node attention for feature fusion; **Only semantic attention for feature fusion.
\end{tablenotes}
\end{threeparttable}
}
\label{ablation}
\end{table}

\subsection{Performance evaluation}
We compare our PerfGAT with other state-of-the-art models (Table~\ref{baseline}). The comparison methods are grouped into three categories: CNN-based spatial model for T1 and rCBV images, non-GNN spatiotemporal models (ResNet+ConvLSTM\cite{yao_deepprognosis_2021},  SwiFT\cite{kim2023swift}), and GNN-based spatiotemporal models (GCN\cite{kipf2017semisupervised}, GIN\cite{xu2019powerful}, GAT\cite{veličković2018graph}, EGAT\cite{kaminski_rossmann-toolbox_2022}, STGCN\cite{yu_spatio-temporal_2018} and BrainGNN\cite{li_braingnn_2021}).
% in terms of Accuracy (ACC), Balanced accuracy (b-ACC), Specificity (SPEC), Sensitivity (SENS), F1 score and Area Under the Curve (AUC). 
The results show that PerfGAT outperforms all the comparison models in most metrics, achieving at least 21.2$\%$, 58.2$\%$ and 16.5$\%$ improvement over other models in balanced accuracy, F1 score and AUC, respectively. Notably, the achievement in superior balanced-accuracy and sensitivity indicating the model performance in learning the minority class. Figure~\ref{ROC} plots the ROC curves of all models. This underscores the effectiveness of our method in classifying IDH mutations even in unbalanced datasets. Note that sensitivities are similar due to limited negative samples in test sets.

\subsection{Ablation Experiments}
Table~\ref{ablation} shows that graph structure learning effectively reduces noise, improving at least 2.7$\%$ in all metrics. Regarding feature fusion, dual-attention mechanisms outperform simple concatenation by 9.8$\%$, 24.9$\%$, 12.6$\%$ and 6.3$\%$ in balanced accuracy, sensitivity, F1 and AUC, respectively. Compared to random resampling, the recombine augmentation method achieved at least 3.6$\%$ improvement in most metrics except sensitivity, indicating less data distortion. Additionally, when removing the spatial graph or local tumor information, the model performance decreases in most metrics, suggesting that informative features can be identified from both local lesions and the global brain.

\section{Conclusion}
In this study, we propose a spatiotemporal GNN framework, PerfGAT, for predicting the genotype of glioma using pMRI data. Our approach optimizes temporal connections using graph structure learning, and fuses spatiotemporal features through a dual-attention mechanism. To tackle the class imbalance challenge, we propose a class-balanced augmentation technique tailored to spatiotemporal features. Our proposed method addresses key challenges in pMRI analysis, demonstrating superior performance compared to state-of-the-art approaches. Future research may include multi-modal graph analysis and clinical validation.

\bibliographystyle{splncs04}
\bibliography{miccai2024_pMRI}

\begin{thebibliography}{10}
\providecommand{\url}[1]{\texttt{#1}}
\providecommand{\urlprefix}{URL }
\providecommand{\doi}[1]{https://doi.org/#1}

\bibitem{Chawla_2002}
Chawla, N.V., Bowyer, K.W., Hall, L.O., Kegelmeyer, W.P.: Smote: Synthetic minority over-sampling technique. Journal of Artificial Intelligence Research  \textbf{16},  321–357 (Jun 2002). \doi{10.1613/jair.953}, \url{http://dx.doi.org/10.1613/jair.953}

\bibitem{cui2019classbalanced}
Cui, Y., Jia, M., Lin, T.Y., Song, Y., Belongie, S.: Class-balanced loss based on effective number of samples (2019)

\bibitem{heo_deep_2023}
Heo, D., Lee, J., Yoo, R.E., Choi, S.H., Kim, T.M., Park, C.K., Park, S.H., Won, J.K., Lee, J.H., Lee, S.T., Choi, K.S., Lee, J.Y., Hwang, I., Kang, K.M., Yun, T.J.: Deep learning based on dynamic susceptibility contrast {MR} imaging for prediction of local progression in adult-type diffuse glioma (grade 4). Sci Rep  \textbf{13}(1),  13864 (Aug 2023). \doi{10.1038/s41598-023-41171-9}, \url{https://www.nature.com/articles/s41598-023-41171-9}, number: 1 Publisher: Nature Publishing Group

\bibitem{kaminski_rossmann-toolbox_2022}
Kamiński, K., Ludwiczak, J., Jasiński, M., Bukala, A., Madaj, R., Szczepaniak, K., Dunin-Horkawicz, S.: Rossmann-toolbox: a deep learning-based protocol for the prediction and design of cofactor specificity in {Rossmann} fold proteins. Brief Bioinform  \textbf{23}(1),  bbab371 (Jan 2022). \doi{10.1093/bib/bbab371}

\bibitem{kang2020decoupling}
Kang, B., Xie, S., Rohrbach, M., Yan, Z., Gordo, A., Feng, J., Kalantidis, Y.: Decoupling representation and classifier for long-tailed recognition (2020)

\bibitem{kim2023swift}
Kim, P.Y., Kwon, J., Joo, S., Bae, S., Lee, D., Jung, Y., Yoo, S., Cha, J., Moon, T.: Swift: Swin 4d fmri transformer (2023)

\bibitem{kingma2017adam}
Kingma, D.P., Ba, J.: Adam: A method for stochastic optimization (2017)

\bibitem{kipf2017semisupervised}
Kipf, T.N., Welling, M.: Semi-supervised classification with graph convolutional networks (2017)

\bibitem{li_decoding_2019}
Li, C., Wang, S., Liu, P., Torheim, T., Boonzaier, N.R., van Dijken, B.R., Schönlieb, C.B., Markowetz, F., Price, S.J.: Decoding the {Interdependence} of {Multiparametric} {Magnetic} {Resonance} {Imaging} to {Reveal} {Patient} {Subgroups} {Correlated} with {Survivals}. Neoplasia  \textbf{21}(5),  442--449 (May 2019). \doi{10.1016/j.neo.2019.03.005}, \url{https://www.sciencedirect.com/science/article/pii/S1476558618306377}

\bibitem{li_braingnn_2021}
Li, X., Zhou, Y., Dvornek, N., Zhang, M., Gao, S., Zhuang, J., Scheinost, D., Staib, L.H., Ventola, P., Duncan, J.S.: {BrainGNN}: {Interpretable} {Brain} {Graph} {Neural} {Network} for {fMRI} {Analysis}. Medical Image Analysis  \textbf{74},  102233 (Dec 2021). \doi{10.1016/j.media.2021.102233}, \url{https://www.sciencedirect.com/science/article/pii/S1361841521002784}

\bibitem{liu2021swin}
Liu, Z., Lin, Y., Cao, Y., Hu, H., Wei, Y., Zhang, Z., Lin, S., Guo, B.: Swin transformer: Hierarchical vision transformer using shifted windows (2021)

\bibitem{louis_2021_2021}
Louis, D.N., Perry, A., Wesseling, P., Brat, D.J., Cree, I.A., Figarella-Branger, D., Hawkins, C., Ng, H.K., Pfister, S.M., Reifenberger, G., Soffietti, R., von Deimling, A., Ellison, D.W.: The 2021 {WHO} {Classification} of {Tumors} of the {Central} {Nervous} {System}: a summary. Neuro Oncol  \textbf{23}(8),  1231--1251 (Jun 2021). \doi{10.1093/neuonc/noab106}, \url{https://www.ncbi.nlm.nih.gov/pmc/articles/PMC8328013/}

\bibitem{Pinto_2018}
Pinto, A., Pereira, S., Meier, R., Alves, V., Wiest, R., Silva, C.A., Reyes, M.: Enhancing Clinical MRI Perfusion Maps with Data-Driven Maps of Complementary Nature for Lesion Outcome Prediction, p. 107–115. Springer International Publishing (2018). \doi{10.1007/978-3-030-00931-1\_13}, \url{http://dx.doi.org/10.1007/978-3-030-00931-1\_13}

\bibitem{rohlfing_sri24_2010}
Rohlfing, T., Zahr, N.M., Sullivan, E.V., Pfefferbaum, A.: The {SRI24} multichannel atlas of normal adult human brain structure. Hum Brain Mapp  \textbf{31}(5),  798--819 (May 2010). \doi{10.1002/hbm.20906}

\bibitem{shi_convolutional_2015}
SHI, X., Chen, Z., Wang, H., Yeung, D.Y., Wong, W.k., WOO, W.c.: Convolutional {LSTM} {Network}: {A} {Machine} {Learning} {Approach} for {Precipitation} {Nowcasting}. In: Advances in {Neural} {Information} {Processing} {Systems}. vol.~28. Curran Associates, Inc. (2015), \url{https://papers.nips.cc/paper\_files/paper/2015/hash/07563a3fe3bbe7e3ba84431ad9d055af\-Abstract.html}

\bibitem{veličković2018graph}
Veličković, P., Cucurull, G., Casanova, A., Romero, A., Liò, P., Bengio, Y.: Graph attention networks (2018)

\bibitem{wang2021heterogeneous}
Wang, X., Ji, H., Shi, C., Wang, B., Cui, P., Yu, P., Ye, Y.: Heterogeneous graph attention network (2021)

\bibitem{wang_multi-task_2023}
Wang, X., Price, S., Li, C.: Multi-task {Learning} of {Histology} and {Molecular} {Markers} for {Classifying} {Diffuse} {Glioma} (Jun 2023). \doi{10.48550/arXiv.2303.14845}, \url{http://arxiv.org/abs/2303.14845}, arXiv:2303.14845 [cs, eess]

\bibitem{wei_multi-modal_2023}
Wei, Y., Chen, X., Zhu, L., Zhang, L., Schönlieb, C.B., Price, S., Li, C.: Multi-{Modal} {Learning} for {Predicting} the {Genotype} of {Glioma}. IEEE Transactions on Medical Imaging  \textbf{42}(11),  3167--3178 (Nov 2023). \doi{10.1109/TMI.2023.3244038}, \url{https://ieeexplore.ieee.org/abstract/document/10042035}, conference Name: IEEE Transactions on Medical Imaging

\bibitem{wei_quantifying_2021}
Wei, Y., Li, C., Price, S.J.: Quantifying {Structural} {Connectivity} in {Brain} {Tumor} {Patients}. In: de~Bruijne, M., Cattin, P.C., Cotin, S., Padoy, N., Speidel, S., Zheng, Y., Essert, C. (eds.) Medical {Image} {Computing} and {Computer} {Assisted} {Intervention} – {MICCAI} 2021. pp. 519--529. Lecture {Notes} in {Computer} {Science}, Springer International Publishing, Cham (2021). \doi{10.1007/978-3-030-87234-2\_49}

\bibitem{xu2019powerful}
Xu, K., Hu, W., Leskovec, J., Jegelka, S.: How powerful are graph neural networks? (2019)

\bibitem{yao_deepprognosis_2021}
Yao, J., Shi, Y., Cao, K., Lu, L., Lu, J., Song, Q., Jin, G., Xiao, J., Hou, Y., Zhang, L.: {DeepPrognosis}: {Preoperative} prediction of pancreatic cancer survival and surgical margin via comprehensive understanding of dynamic contrast-enhanced {CT} imaging and tumor-vascular contact parsing. Medical Image Analysis  \textbf{73},  102150 (Oct 2021). \doi{10.1016/j.media.2021.102150}, \url{https://www.sciencedirect.com/science/article/pii/S1361841521001961}

\bibitem{yu_spatio-temporal_2018}
Yu, B., Yin, H., Zhu, Z.: Spatio-{Temporal} {Graph} {Convolutional} {Networks}: {A} {Deep} {Learning} {Framework} for {Traffic} {Forecasting}. In: Proceedings of the {Twenty}-{Seventh} {International} {Joint} {Conference} on {Artificial} {Intelligence}. pp. 3634--3640 (Jul 2018). \doi{10.24963/ijcai.2018/505}, \url{http://arxiv.org/abs/1709.04875}, arXiv:1709.04875 [cs, stat]

\bibitem{zhang_brainusl_2023}
Zhang, P., Wen, G., Cao, P., Yang, J., Zhang, J., Zhang, X., Zhu, X., Zaiane, O.R., Wang, F.: {BrainUSL}: {Unsupervised} {Graph} {Structure} {Learning} for {Functional} {Brain} {Network} {Analysis}. In: Greenspan, H., Madabhushi, A., Mousavi, P., Salcudean, S., Duncan, J., Syeda-Mahmood, T., Taylor, R. (eds.) Medical {Image} {Computing} and {Computer} {Assisted} {Intervention} – {MICCAI} 2023. pp. 205--214. Lecture {Notes} in {Computer} {Science}, Springer Nature Switzerland, Cham (2023). \doi{10.1007/978-3-031-43993-3\_20}

\end{thebibliography}

\end{document}